\documentclass[a4paper]{article}

\usepackage{INTERSPEECH2022}
\usepackage[colorinlistoftodos,prependcaption]{todonotes}
\usepackage{makecell}
\usepackage{url}
\usepackage{amsmath}
\usepackage{color}
\usepackage[hidelinks]{hyperref}
\usepackage{wrapfig}
\usepackage{graphicx}
\usepackage{caption}
\usepackage{subcaption}
\usepackage{enumitem}
\usepackage{scrextend}
\DeclareMathOperator*{\argmax}{arg\,max}
\makeatletter
\newcommand{\printfnsymbol}[1]{%
  \textsuperscript{\@fnsymbol{#1}}%
}
\makeatother
\title{AdvEst: Adversarial Perturbation Estimation to Classify and Detect Adversarial Attacks against Speaker Identification}
\name{Sonal Joshi\printfnsymbol{1} \thanks{\printfnsymbol{1}equal contribution},
        Saurabh Kataria\printfnsymbol{1},
        Jes{\'u}s Villalba\printfnsymbol{1},
        Najim Dehak}

\address{Center for Language and Speech Processing, 
Johns Hopkins University, Baltimore, MD}
\email{\{sjoshi12,skatari1,jvillal7,ndehak3\}@jhu.edu}
\begin{document}
\maketitle
\begin{abstract}
Adversarial attacks pose a severe security threat to the state-of-the-art speaker identification systems, thereby making it vital to propose countermeasures against them. Building on our previous work that used representation learning to classify and detect adversarial attacks, we propose an improvement to it using AdvEst, a method to estimate adversarial perturbation. First, we prove our claim that training the representation learning network using adversarial perturbations as opposed to adversarial examples (consisting of the combination of clean signal and adversarial perturbation) is beneficial because it eliminates nuisance information. At inference time, we use a time-domain denoiser to estimate the adversarial perturbations from adversarial examples. Using our improved representation learning approach to obtain attack embeddings (\emph{signatures}), we evaluate their performance for three applications: known attack classification, attack verification, and unknown attack detection. We show that common attacks in the literature (Fast Gradient Sign Method (FGSM), Projected Gradient Descent (PGD), Carlini-Wagner (CW) with different $L_p$ threat models) can be classified with an accuracy of $\sim96$\%. We also detect unknown attacks with an equal error rate (EER) of $\sim$9\%, which is absolute improvement of $\sim$12\% from our previous work.
\end{abstract}

\section{Introduction}
\label{sec:intro}


Adversarial examples are malicious inputs designed to make deep neural networks produce incorrect prediction~\cite{szegedy-iclr14}. They are generated by adding carefully computed perturbation to clean signals (known by the term \textit{benign} in the literature). Mathematically, $ \mathbf{x} + \mathbf{\delta} = \mathbf{x'}$, where $\mathbf{\delta}$ denotes the adversarial perturbation\footnote{\label{note1}For detailed explanation, please refer to Section~\ref{sec:adv_attacks}} being added to the benign signal $\mathbf{x}$ to obtain adversarial example $\mathbf{x'}$. 
These adversarial examples are (almost) inaudible to humans and pose a severe security threat to state-of-the-art speaker recognition systems. To further exacerbate the problem, several adversarial attack algorithms have been proposed~\cite{kreuk-icassp18,Xie2020,wang2020inaudible,abdullahsok2020} and new attack algorithms and variants come out frequently. This makes proposing practical countermeasures against these attacks of utmost importance. 

Most of the countermeasures in the literature can be primarily classified into two types--proactive and reactive~\cite{wang2020adversarial,joshi2021study}. Proactive countermeasures focus on making the system under attack (a.k.a. the victim network) inherently more robust to adversarial examples. This is usually done by re-training the system model, say by adversarial training or fine-tuning~\cite{madry2018towards}. The limitation of such defenses is that they may be susceptible to adversarial attack algorithms that they are not trained for~\cite{zhang2019limitations}. On the other hand, reactive countermeasures focus on having an additional front-end block that either pre-processes the input or does \emph{denial of service}, say using an adversarial attack detector, that acts as a gate in order to reject any incoming adversarial examples~\cite{rajaratnam2018noise,li2020investigating,yang2018characterizing}. However, the system becomes less user-friendly if the detector has a high false alarm rate. Hence, it becomes vital to have a system that can accurately classify the types of adversarial attacks and detect the presence of an unknown attack so that the best possible defense can be selected accordingly. An additional benefit of such classification/detection countermeasure is that it may help identify if two attacks are from the same attacker or different. This is particularly important in forensics applications. However, this area of research is still in the nascent stage, with few published works in image and text classification~\cite{moayeri2021sample,xie2022identifying}.

We focus on attacks on speaker recognition systems, particularly on the state-of-the-art x-vector based system. Our previous work~\cite{villalba2021representation} proposed to use embeddings obtained by representation learning of adversarial examples as \emph{attack signatures} to retrieve information about the adversarial attack. This information includes attack algorithm type and threat model\footref{note1}, Signal to Adversarial noise ratio, etc. could help in knowing about attacker identity and intention. Previous work~\cite{villalba2021representation} shows that representation learning networks trained on speaker identification attacks are transferable to speaker verification attacks. Hence, we focus  on speaker identification tasks in this work. 

In~\cite{villalba2021representation}, we use signature extractors that take the adversarial signal $\mathbf{x'}$ as input.
However, $\mathbf{x'}$ consists of both benign signal $\mathbf{x}$ and adversarial perturbation $\mathbf{\delta}$. We argue that, instead of $\mathbf{x'}$, $\mathbf{\delta}$ should have more relevant information about the attack.  We verify this claim--that using adversarial perturbation $\mathbf{\delta}$ instead of $\mathbf{x'}$ is better for the system--empirically. 
Note that, at training time, we can generate of many attacks and, therefore, have access to benign and adversarial example pair $\mathbf{(x',x)}$, it is straightforward to obtain $\mathbf{\delta}$ by simply subtracting  $\mathbf{x}$ from $\mathbf{x'}$. 
However, at inference time, obtaining $\mathbf{\delta}_\text{inference}$ is not trivial due to two major challenges. First, we only know the adversarial example $\mathbf{\mathbf{x}'_\text{inference}}$ and do not have access to  $\mathbf{\mathbf{x}_\text{inference}}$, the corresponding benign signal. Second, although one may argue that this can be estimated using a \textit{simple} speech enhancement/denoising network trained using $\mathbf{(x',x)}$ pairs available for training adversarial examples, with $\mathbf{x}$ being target for the input $\mathbf{x'}$; it is not trivial to obtain a good network that can accurately estimate benign signal $\mathbf{x}_{inference}$ and hence  $\mathbf{\delta}_{inference}$. This challenge is due to the fact that the current speech enhancement/denoising networks are optimized for environmental noises which are very different from adversarial perturbations~\cite{shamir2021dimpled}. We propose a solution to both these problems by proposing a novel method using combination of representation learning and adversarial perturbation estimation to detect and classify adversarial attacks on speaker identification (SID) network. Using our proposed method for adversarial perturbation estimation (AdvEst), we evaluate three applications of the signature extractor as reactive countermeasures to mitigate the threat of adversarial attacks:
\begin{itemize}[leftmargin=*]
\item \textbf{Task 1 - Known attack classification:} Given an adversarial example, the goal is to classify an attack into one of the known attack classes. 
\item  \textbf{Task 2 - Attack verification}: Given two adversarial examples, the goal is to identify if the attack belong to same or different known attack class. This task is similar to speaker verification, but for attacks.
\item \textbf{Task 3 - Unknown attack detection }: Given an adversarial example, the goal is to identify if they belong to an already known class or a novel class.
\end{itemize}

We briefly introduce adversarial attacks on speaker identification in Section~\ref{sec:adv_attacks} followed by outline of the proposed method in Section~\ref{sec:proposed}. The experiments and corresponding results with discussions are in Section~\ref{sec:exp} and~\ref{sec:results}, while the conclusions are in Section~\ref{sec:conlusion}. 

\section{Adversarial Attacks on Speaker Identification}
\label{sec:adv_attacks}
The goal of speaker identification is to predict the identity of the speaker amongst a given pool of known speakers. 

The speaker identification network $H_\theta(\mathbf{x})$ returns the posterior probabilities for the speaker classes.
An adversary attacks the clean signal $\mathbf{x}$ by adding an adversarial perturbation $\mathbf{\delta}$ 

that solves the optimization problem,
\begin{equation}
  \delta = \argmax_{\delta'\in\Delta} \ell(H_\theta(\mathbf{x} + \mathbf{\delta'}),y) \hspace{0.5cm} s.t. \hspace{0.5cm} y \neq y',
\end{equation}
 where, $\Delta$ represents the set of allowable perturbations, $y$ is the ground truth speaker label and $\ell(\cdot)$ is the loss function--e.g., categorical cross-entorpy. $\mathbf{x'} = \mathbf{x} + \mathbf{\delta}$ is the adversarial example and  $y'$ is the speaker label predicted from $\mathbf{x'}$. 

\subsection{Threat Model}
To keep the adversarial perturbation imperceptible, we impose constraints on $\Delta$. Most attacks enforce $ \lVert \mathbf{\delta} \rVert_p\leq \epsilon$, where the $\lVert \cdot \rVert_p$ denotes $L_p$ norm and $\epsilon$ is a scalar, which bounds the strength of the attack. The lower the $\epsilon$, the more imperceptible the attack. The choice of the norm type and bound is called as \textit{threat model}. Common threat models are $L_0$, $L_2$ and $L_\infty$.

\subsection{Attack Algorithms}

The adversarial perturbation $\mathbf{\delta}$ can be optimized by different algorithms. Common methods are Fast Gradient Sign Method (FGSM)~\cite{Goodfellow2015}, Projected Gradient Descent (PGD)~\cite{madry2018towards}, Carlini-Wagner (CW)~\cite{carlini-16}. For detailed working of these adversarial attack algorithms, the reader is encouraged to read \cite{villalba2021representation} and \cite{joshi2021study}.

\subsection{Attack Classification}
Attacks can be classified based on threat model $\{L_0, L_2, L_\infty\}$, algorithm \{FGSM, PGD, CW\} or combination of algorithm and threat model \{PGD-$L_0$, PGD-$L_2$, PGD-$L_\infty$\}. In this work, we focus on the latter as it is a more challenging task.

\section{Proposed Method}
\label{sec:proposed}

\subsection{Signature Extractor Network}

We used an x-vector style network~\cite{snyder-icassp18} to extract embeddings, i.e. attack signatures. Henceforth, this network is termed as signature extractor network. 

The signature extractor was composed of Thin-ResNet34 encoder, statistics pooling and classification head with additive angular margin softmax (AAM-softmax), similar to our previous work~\cite{villalba2021representation}.
In our previous work, we train the signature extractor directly on adversarial examples $\mathbf{x'}$. However, we claim that the job of signature extractor will be easier if it is given $\mathbf{\delta}$ instead, as the benign signal does not contain any information related to the adversarial attack type\footnote{We prove this claim empirically later in Section \ref{sec:results}}. One issue is that, although we can have $\mathbf{\delta}$ available for training data $\mathbf{\delta} =\mathbf{x'}- \mathbf{x}$, henceforth called oracle adversarial perturbation, it cannot be computed at inference time as $\mathbf{x}$ is not available. Hence, we propose to use an Adversarial Perturbation Estimation (AdvEst) Network to obtain $\mathbf{\hat{\delta}}=\mathbf{x'}- \mathbf{\hat{x}}$ where $\mathbf{\hat{x}}$ is the benign signal predicted by AdvEst and $\mathbf{\hat{\delta}}$ is the estimated adversarial perturbation.

\subsection{Adversarial Perturbation Estimation Network}
\label{ssec:cgan}
The Adversarial Perturbation Estimation (AdvEst) network is a denoiser model that is trained to predict the benign signal given the adversarial input, and, then, estimates the perturbation as the difference between the adversarial and predicted benign signals.
We require AdvEst model to work in time-domain and able to predict unnatural noises like adversarial perturbation. However, it was not trivial to find denoiser that could estimate adversarial perturbation because most denoiser models and loss functions in speech enhancement literature are geared towards predicting natural environmental noises.
We experimented with different deep regression and generative models with a variety of loss functions like Multi-Resolution Short-time Fourier Transform (MRSTFT) loss, $L_1$ loss, and $L_2$ loss for known attack classification task. Our findings indicated the superiority of the conditional GAN (CGAN) generative model~\cite{mirza2014conditional,goodfellow2014generative} over conventional regression models. 

CGAN is a supervised variant of GAN which maps signals from one \emph{domain} to the another by learning to sample from the respective conditional distribution.
CGAN consists of two networks: generator $G$ (or $G_{\mathcal{A}\rightarrow\mathcal{B}}$) and discriminator $D$ (or $D_{\mathcal{B}}$).
$D$ learns to differentiate between the real benign signals and the ones generated by $G$.
$\mathcal{G}$ learns to map adversarial signals (domain $\mathcal{A}$) to their benign counterpart (domain $\mathcal{B}$) by ``fooling'' $\mathcal{D}$ \cite{goodfellow2014generative}.
Minimax formulation is 
\begin{equation}
\label{eq:cgan}
\begin{split}
    \min_{\mathcal{G}_{\mathcal{A}\rightarrow\mathcal{B}}} \max_{\mathcal{D}_{\mathcal{B}}} \mathcal{L}_{\text{CGAN}}, \hspace{1em} \text{where} \hspace{1em}
        \mathcal{L}_{\text{CGAN}} = \mathcal{L}_{\text{adv}} + \lambda_{\text{sup}}\mathcal{L}_{\text{sup}}.
\end{split}
\end{equation}
The objective function $\mathcal{L}_{\text{CGAN}}$ consists of two terms: regression loss ($\mathcal{L}_{\text{sup}}$) and adversarial loss ($\mathcal{L}_{\text{adv}}$).
The hyper-parameter $\lambda_{\text{sup}}$ is the weight for the supervision/regression loss.
The adversarial loss $\mathcal{L}_{\text{adv}}$ is based on Dual Contrastive Loss (DCL)~\cite{yu2021dual}:
\begin{align}
\label{eq:adv}
    &\mathcal{L}_{\text{adv}} = -\mathbb{E}_{\mathbf{x}\sim P_{\mathcal{B}}}[\log (1 + \sum_{\mathbf{x'}\sim P_{\mathcal{A}}} \exp({D(G(\mathbf{x'})) - D(\mathbf{x})}))]
    \notag\\
    &\hspace{1.1em}-\mathbb{E}_{\mathbf{x'}\sim P_{\mathcal{A}}}[\log (1 + \sum_{\mathbf{x}\sim P_{\mathcal{B}}} \exp({D(G(\mathbf{x'})) - D(\mathbf{x})}))],
\end{align}
where $D$ returns logit of the posterior probability for the real input.
Here, $\mathbf{x}$ and $\mathbf{x'}$ are pre-computed (Sec.~\ref{sec:attack_gen}) corresponding pairs of benign and adversarial signals, whose marginal and joint distributions are referred to as \{$P_{\mathcal{B}}$,$P_{\mathcal{A}}$\} and $P_{\mathcal{B,A}}$ respectively.
The supervision loss $\mathcal{L}_{\text{sup}}$ is MRSTFT auxiliary loss whose details which can be found in \cite{yamamoto2020parallel}.

The adversarial perturbation in a given adversarial signal $\mathbf{x'}$ is estimated as:
\begin{equation}
    \mathbf{\hat{\delta}} = \mathbf{x'} - \mathbf{\hat{x}}, \hspace{0.5em} \text{where} \hspace{0.5em} \mathbf{\hat{x}} = G_{\mathcal{A}\rightarrow\mathcal{B}}(\mathbf{x'}).
\end{equation}

\section{Experimental Setup}
\label{sec:exp}

\subsection{Speaker Identification}

All experiments are performed using VoxCeleb2 dataset~\cite{Nagrani2020}. For speaker identification task, we need train and test speakers to be the same. Therefore, we used the
\textit{VoxCeleb2-dev} set and split it into two parts in the ratio 90\%-10\% per speaker. The larger set, henceforth called \textit{VoxCeleb2-dev-train} is used to train the victim network, while the smaller set, henceforth called \textit{VoxCeleb2-dev-test} is used as the test set for the speaker identification task. This experimental setup is same as our previous work~\cite{villalba2021representation}. x-Vector architecture for the victim network was based on a Thin-ResNet34 similar to the ones used in~\cite{zeinali2019but,Villalba2020a}, with 256 embedding dimension, additive angular margin of $m=0.3$. This network in clean/benign condition gives  1.94\%, 1.91\%, and 3.2\% EER for VoxCeleb1 original, entire and hard tasks respectively. We refrain from using a larger x-vector network since a larger network would require more time for generating adversarial attacks. 

\subsection{Adversarial attack dataset}
\label{sec:attack_gen}
With the victim network trained, we generate attacks using various algorithms and threat models commonly used in the literature--Fast Gradient Sign Method (FGSM), Iterative FGSM (Iter-FGSM), Projected Gradient Descent (PGD-$L_\infty$/$L_1$/$L_2$) and Carlini \& Wagner (CW-$L_\infty$/$L_0$/$L_2$) attacks. In order to cover the maximum subspace of attacks, we used a wide variety of attack hyper-parameters by sampling from different probability distributions as done in our previous work~\cite{villalba2021representation}. To obtain a balanced set, for every algorithm-threat model class, we randomly sample 25\% of utterances in the \textit{VoxCeleb2-dev-train} set and use it for attack generation. If a generated attack is successful in making the speaker identification network predict incorrect speaker, we add it to our database. In other words, our adversarial attack dataset contains only successful attacks. 

The generated attack dataset is quite large and has 190k utterances corresponding to \textit{VoxCeleb2-dev-train} and 10.5k  corresponding to \textit{VoxCeleb2-dev-test} sets. We use a recipe from the open source toolkit Hyperion to generate these attacks. We split the data into two sets-known attacks and unknown attacks. Unknown attacks set contains the Carlini \& Wagner (CW-$L_\infty$/$L_0$/$L_2$) attacks, while the remainder are put in the known attacks set. The choice of CW attacks for unknown was considering the fact that they were the most stealth attacks and less common as compared to PGD.
For more details, the reader is encouraged to refer to our previous work~\cite{villalba2021representation}. 

\subsection{Adversarial Perturbation Estimation Network}
\label{sec:cgan_train}
We use time-domain architectures for the generator and discriminator for our CGAN based denoiser.
For generator, we choose Conv-TasNet (or simply TasNet)~\cite{luo2019conv}, which is a Convolutional Neural Network (CNN) consisting of three stages: \emph{encoder}, \emph{separator}, and \emph{decoder}.
The \emph{encoder} and \emph{decoder} are single 1-D convolutional layers, while \emph{separator} stage consists of multiple 1-D CNNs (termed \emph{stacks}) whose individual outputs are combined to produce \emph{masks} which are multiplied element-wise to the input mixture and passed on to the \emph{decoder} stage.
We choose number of stacks as 1, number of layers per stack as 6, and the number of output channels (decoder) as 1.
The number of channels in encoder is 128, kernel size is 16 (1 ms as per sampling frequency of 16 KHz), and stride is 8.
The number of input and output channels in separator are 128 and 1024 respectively, and dilation is exponentially increasing with a factor of 2.
Total number of parameters are 1.6M and receptive field is 511.
For more details, refer \cite{luo2019conv}.
The discriminator is based on Parallel WaveGAN~\cite{yamamoto2020parallel}.
It is a deep 1-D CNN with 10 convolutional layers with 80 channels, kernel size of 3, stride of 1, and linearly increasing dilation starting from 1.
An additional sigmoid layer is appended at the end of the discriminator for added training stability.

CGAN is trained with Alternating Gradient Descent (AGD) algorithm~\cite{goodfellow2014generative} with the generator updating at twice the frequency of the discriminator.
$\lambda_{\text{sup}}$ is set to 0.001.
For training the denoiser, we use \emph{voxceleb2-dev-train} set explained above.
We remove silence from it using energy-based Voice Activity Detection (VAD)~\cite{villalba2021representation}. The learning rates of generator and discriminator are 0.0002 and 0.0001 respectively, both decreasing linearly at each step till 1e-8 and the optimizer is Adam~\cite{kingma2014adam} with betas=(0.5,0.999).
For supervised learning, the noisy counterpart of the clean training data is needed, for which we used the attacks described in Section~\ref{sec:attack_gen}. We train two networks --one using all attacks (which is used for known attack classification) and second leaving out CW attack (which is used for unknown attack detection experiments). 

\begin{figure*}
\begin{tabular}{cc}
Overall classification accuracy = 92.23\% & Overall classification accuracy = 96.30\%  \\
  \includegraphics[trim=0.5cm 0cm 3.9cm 1.2cm,clip=true,width=0.5\linewidth]{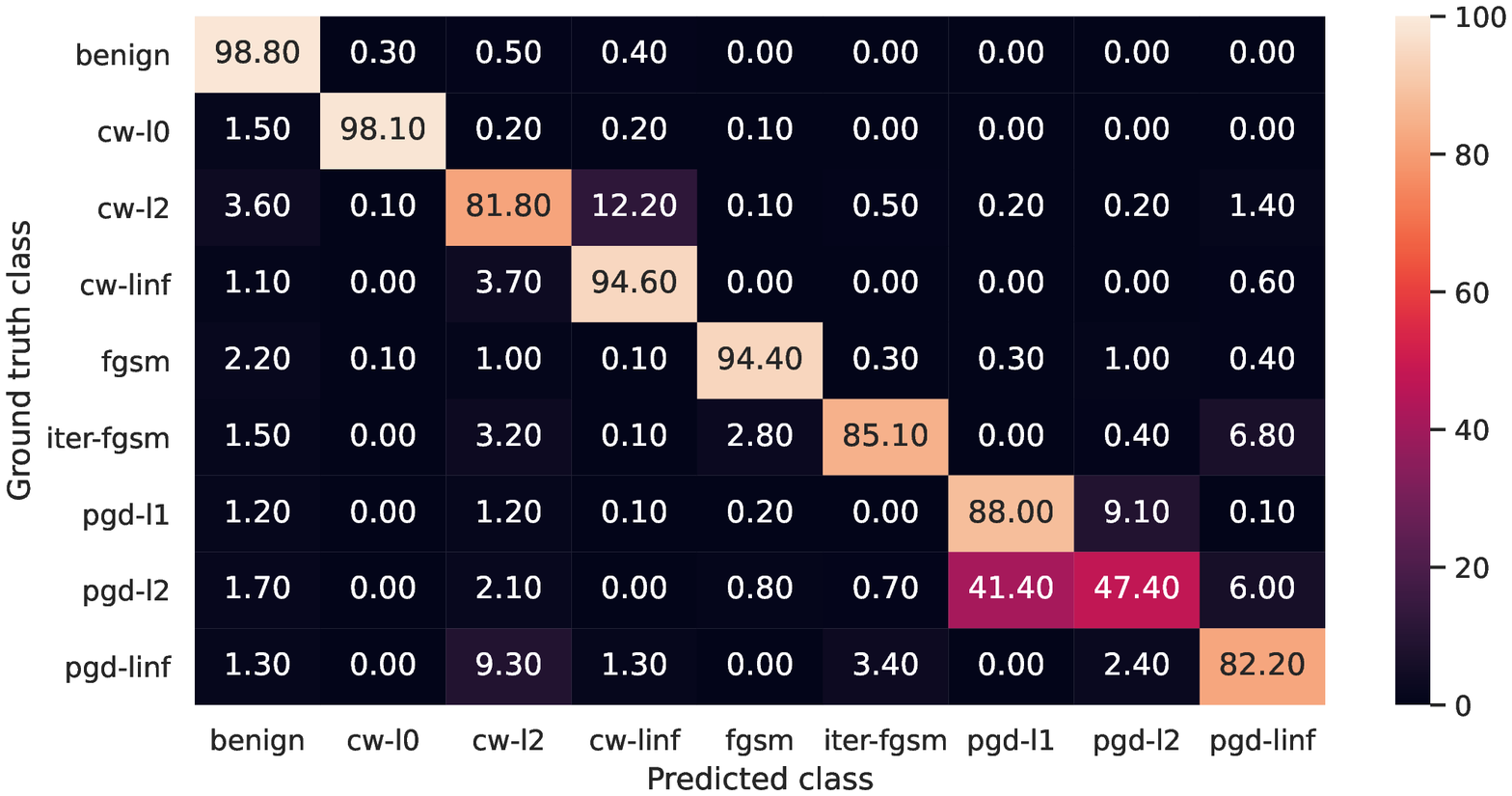} &   \includegraphics[trim=0.5cm 0cm 3.9cm 1.2cm,clip=true,width=0.5\linewidth]{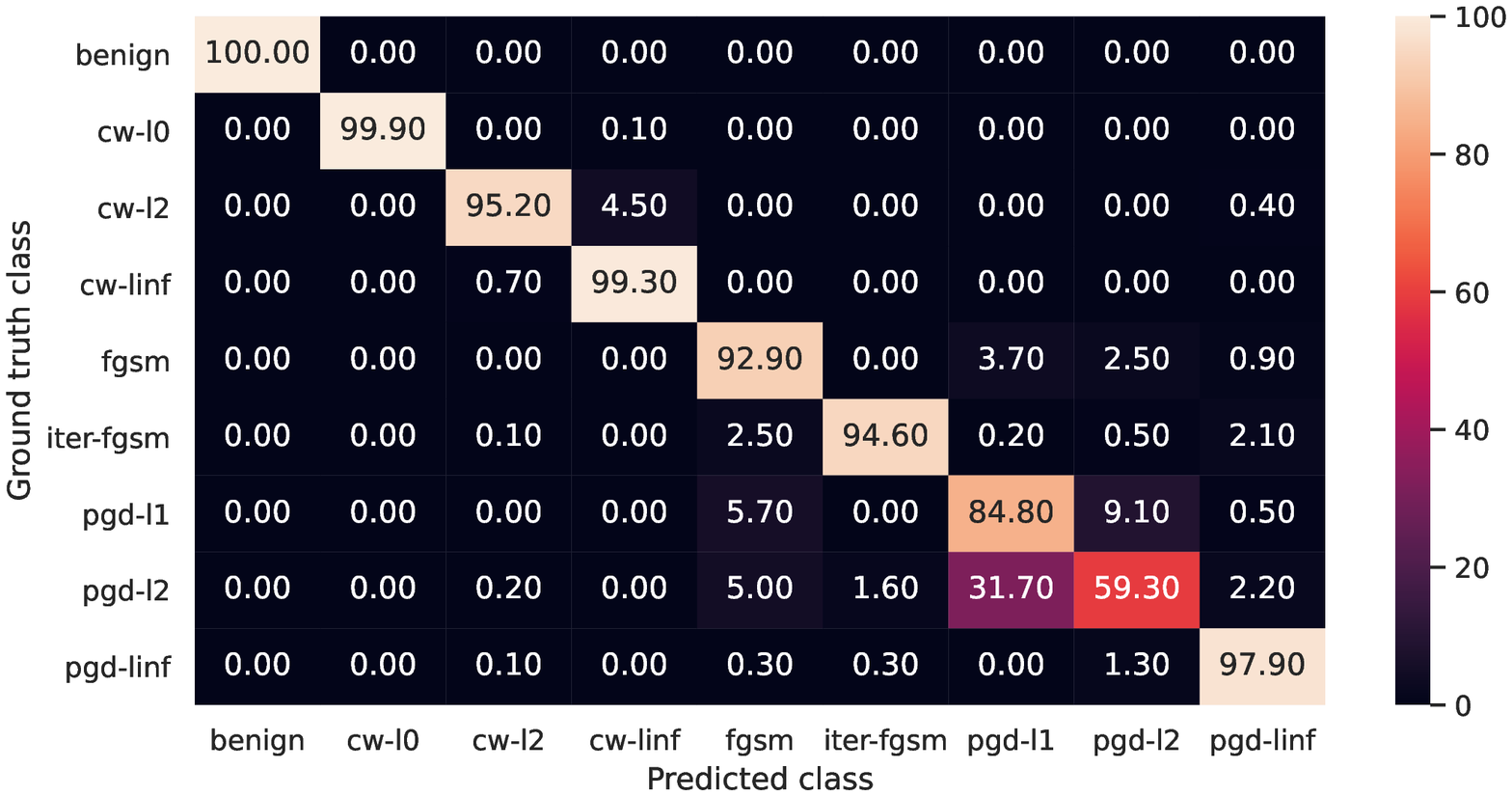} \\
  (a) Baseline. & (b) Proposed.
\end{tabular}
\vspace{-3mm}
\caption{Confusion matrices for known attack classification using different methods.}
\vspace{-3mm}
\label{fig:cm}
\end{figure*}

\begin{table}
  \caption{Experimental setup for signatures extractor networks}
  \vspace{-3mm}
  \label{tab:exps}
  \begin{tabular}{@{}lcc@{}}
    \toprule
    \textbf{Experiment Name\hspace{3cm}} & \textbf{Train} & \textbf{Test} \\
    Description\\
    \midrule
    \textbf{Baseline} & $\mathbf{x'}$ & $\mathbf{x'}$   \\
    \multicolumn{3}{l}{Train and test using adversarial example} \\
    \textbf{Oracle Perturbation} & $\delta$ & $\delta$ \\
    \multicolumn{3}{l}{Train and test using Oracle Adversarial Perturbation} \\
    \textbf{Estimated Adversarial Perturbation} & $\hat{\delta}$ & $\hat{\delta}$  \\
    \multicolumn{3}{l}{Train and test using Estimated Adversarial Perturbation} \\
    \textbf{Proposed} & $\delta$ & $\hat{\delta}$ \\
    \multicolumn{3}{l}{\makecell[l]{Train using Oracle Adversarial Perturbation and \\ test using Estimated Adversarial Perturbation}} \\
    \bottomrule
  \end{tabular}
  \vspace{-3mm}
\end{table}

\subsection{Systems and Evaluation Tasks}
We evaluate four signature extractor systems as outlined in Table~\ref{tab:exps}. \textbf{Baseline} system utilizes adversarial example ($\mathbf{x'}$ ) at both train and evaluation time, while \textbf{Oracle Perturbation} system utilizes oracle adversarial perturbation ($\delta$). Please note that this is an oracle experiment carried out just to  validate our claim that adversarial perturbations  are better than adversarial examples for training signature extractors. \textbf{Estimated Adversarial Perturbation} system uses adversarial perturbation estimated using AdvEst for both train and inference. Lastly, we have the \textbf{Proposed} system which is trained using oracle adversarial perturbation ($\delta$) while  adversarial perturbation estimated using AdvEst ($\hat{\delta}$) is used at inference time.
We evaluate the performance of the four systems using three tasks as described in Section~\ref{sec:intro}--known attack detection, attack verification, unknown attack detection in Table \ref{tab:exps}.

\vspace{-3mm}
\section{Results and Discussion}
\label{sec:results}
\begin{table}
  \caption{Classification Accuracy (\%) for known attack detection}
  \vspace{-3mm}
  \label{tab:known_acc}
  \centering
  \begin{tabular}{@{}lrr@{}}
    \toprule
    \textbf{System} & \textbf{Accuracy(\%)} \\ 
    \midrule
    \textbf{Baseline} & 92.23 \\
    \textbf{Oracle Perturbation} & 96.26 \\
    \textbf{Estimated Perturbation} &  90.79 \\
    \textbf{Proposed} & \textbf{96.30} \\
    \bottomrule
  \end{tabular}
  \vspace{-3mm}
\end{table}
\textbf{Task 1 - Known attack classification:}
Table~\ref{tab:known_acc} shows the results for known attack classification.
The baseline classification accuracy is 92.23\%, while using Oracle  Perturbation ($\delta$) improves it to  96.26\%. This suggests that removing the benign signal makes the job of signature extractor easier. Using Estimated Adversarial Perturbation, we obtain classification accuracy of 90.79\%. This suggests that AdvEst network predicted adversarial perturbation is worse than oracle. We propose to mitigate this by oracle perturbation while training and using estimated perturbation at inference time. The proposed method yields an overall classification accuracy is 96.30\% which is close to our oracle experiment. Comparing the confusion matrices between baseline and proposed in Figure \ref{fig:cm}(a) and (b), we observe that the accuracies along diagonal improve for all attacks except PGD-$L_1$. Although, we still have problems distinguishing between PGD-$L_1$ and PGD-$L_2$, the percentage of PGD-$L_2$ attacks classified as PGD-$L_1$ is significantly reduced. Overall, all attacks except PGD-$L_1,L_2$ are better than 92\%. Absolute relative improvement range is between -1.5\% to 15.7\%.

\begin{table}
    \caption{EER (\%) for attack verification. The best results are in \textbf{bold} while the second best results are \underline{underlined}}
    \label{tab:verif}
    \vspace{-3mm}
    \centering
    \resizebox{0.9\columnwidth}{!}{
    \begin{tabular}{@{}lrrr@{}}
    \toprule
    \textbf{System} & \multicolumn{3}{c}{\textbf{Attack group EER(\%)}}\\
    \cmidrule{2-4}
    & \textbf{Known} & \textbf{Unknown} & \textbf{\makecell{Known\\+Unknown}}\\
    \midrule
    \textbf{Baseline} & 8.88 & 36.53 & 20.23 \\
    \textbf{Oracle Perturbation} & \textbf{5.03} & \textbf{26.17} & \underline{14.82} \\
    \textbf{Estimated Perturbation} & \underline{5.14} & 40.77 & 21.66 \\
    \textbf{Proposed} & \underline{5.14} & \underline{29.01} & \textbf{14.57}\\
\bottomrule
\end{tabular}}
\vspace{-4mm}
\end{table}
\textbf{Task 2 - Attack verification:}
We consider three cases--known attacks only, unknown attacks only and combination of known and unknown attacks. Table \ref{tab:verif} shows the results for the three conditions using equal error rate (EER) (\%) as the metric. Oracle Perturbation method clearly outperforms the baseline method in all three conditions. The Estimated Perturbation method is almost on-par for Known attacks condition, however is quite worse than Oracle Perturbation method when the conditions have unknown attacks. The Proposed method performs the best for these conditions. Considering that Oracle Perturbation method is an impractical method (exact adversarial perturbation cannot be computed), 
the proposed method is the best method for all conditions.\\

\begin{table}
  \caption{EER (\%) for unknown attack detection. Benign, FGSM and PGD are known attacks, CW attacks are unknown.}
  \label{tab:unknown_eer}
  \vspace{-3mm}
  \centering
   \resizebox{0.9\columnwidth}{!}{
  \begin{tabular}{@{}lrr@{}}
    \toprule
    \textbf{System} & \textbf{With benign} & \textbf{Without benign}\\
    \midrule
    \textbf{Baseline} & 42.86 & 21.36 \\
    \textbf{Oracle Perturbation} & 39.43 & 17.04 \\
    \textbf{Estimated Perturbation}  & 41.45 & 20.75 \\
    \textbf{Proposed} & \textbf{37.99} & \textbf{9.06} \\
    \bottomrule
  \end{tabular}
  \vspace{-6mm}}
\end{table}
\raggedbottom
\textbf{Task 3 - Unknown attack detection:}
Our previous work showed that unknown attacks are easily confused with benign, and hence we consider two cases: with and without benign class. For \emph{without benign} class, we assume that we have an oracle detector telling that system the under attack and the goal detector is to just tell if the attack is known or unknown  detect. We use equal error rate (EER)\% as metric, and the results are shown in Table \ref{tab:unknown_eer}. We can see that using Oracle Perturbation is better than using the adversarial example Baseline. Using Estimated Perturbation, we are not able to train the signature extractor as well as with Oracle Perturbation, and this is indicated by a deteriorated performance (EER increases by $\sim 4$\%). The Proposed method yields 9\% EER. We speculate that this might be due to the widely known limitation of deep neural network-based denoisers--that it achieves good inference-time performance for noise seen during training, but not so good performance for unseen noises~\cite{joshi2018enhanced}. This (undesirable) property seems to be helping with unknown attack detection. We speculate that the denoiser maps the unknown attacks to a different subspace than known attacks, thus making the task of unknown attack detection easier. We are currently investigating this end.
\vspace{-3mm}
\section{Conclusions}
\label{sec:conlusion}
We verify the hypothesis that using just the adversarial perturbation makes the job of learning signatures easier. At inference time, we use AdvEst, an adversarial perturbation estimation network, to help to remove the clean signal from adversarial example. Using our proposed approach, we show that common attacks in the literature can be classified with accuracy as high as $\sim$96\%, while unknown attacks can be detected with an equal error rate (EER) of $\sim$9\%. We are currently performing additional experiments to investigate the reason for performance improvement using AdvEst for unseen attacks.
\vspace{-3mm}
\section{Acknowledgements}
This research has been supported by DARPA RED under contract HR00112090132 \footnote{https://www.darpa.mil/program/reverse-engineering-of-deceptions}

\bibliographystyle{IEEEtran}

\bibliography{mybib}

\end{document}